\documentclass[final,1p,time]{elsarticle}
\usepackage{xfrac}

\usepackage{longtable}
\usepackage{graphicx}

\usepackage{amssymb}
\usepackage{amsmath}

\usepackage{color}

\journal{Journal of Computational Science}

\begin{document}
	\begin{frontmatter}
		
		\title{Towards Exascale Design of Soft Mesoscale Materials}
		
		\author[iit,harv,iac]{Sauro Succi\corref{cor1}}
		\cortext[cor1]{Corresponding author}
		\ead{sauro.succi@iit.it }
		
		\author[cineca]{Giorgio Amati}
		\author[iit]{Fabio Bonaccorso}
		
		\author[iac]{Marco Lauricella, M. Bernaschi}
		
		\author[iac]{Andrea Montessori}
		\author[iit,iac]{Adriano Tiribocchi}

		\address[iit]{Center for Life Nano Science$@$La Sapienza, Istituto Italiano di Tecnologia, 00161 Roma, Italy}
		\address[iac]{Istituto per le Applicazioni del Calcolo CNR, Via dei Taurini 19, 00185 Rome, Italy}
		\address[cineca]{SCAI, SuperComputing Applications and Innovation Department, CINECA, Via dei Tizii, 6, Rome 00185, Italy}
		\address[harv]{John A. Paulson School of Engineering and Applied Sciences, Harvard University, 33 Oxford St., Cambridge, MA 02138, USA}
	\end{frontmatter}
	
	%% main tex
	
\section{Abstract}\label{Abs}
We provide a brief survey of our current developments in the simulation-based 
design of novel families of mesoscale porous materials using computational kinetic theory. 
Prospective applications on exascale computers are also briefly discussed and commented on, with 
reference to two specific examples of soft mesoscale materials: microfluid crystals 
and bi-continuous jels. 
	
\section{Introduction} \label{Introd}

Complex fluid-interface dynamics \cite{fernandez2016fluids,piazza2011soft,doi2013soft,stratford2005colloidal}, disordered liquid-liquid 
emulsions \cite{frijters2012effects,costantini2014highly}, soft-flowing microfluidic crystals \cite{raven2009microfluidic,marmottant2009microfluidics,montessori2019mesoscale, montessori2019modeling},
all stand as complex states of matter which, besides raising new challenges to 
modern non-equilibrium thermodynamics, pave the way to many engineering applications, such as combustion  
and food processing \cite{cant2008introduction,muschiolik2017double}, as well as to questions in fundamental biological
and physiological processes, like blood flows and protein dynamics \cite{bernaschi2019mesoscopic}.

In particular this novel state of soft matter opens up exciting prospects for the design 
of new materials whose effective building blocks are droplets instead of molecules \cite{garstecki2006flowing,raven2009microfluidic,abate2009high,beatus2012physics}.
 
The design of new materials has traditionally provided a relentless stimulus to the development of computational
schemes spanning the full spectrum of scales, from electrons to atoms and molecules, to supramolecular 
structures all the way up to the continuum, encompassing over ten decades in space (say from Angstroms to meters)
and at least fifteen in time (from femtoseconds to seconds, just to fix ideas).
Of course, no single computational model can handle such huge spectrum of scales, each region
being treated by dedicated and suitable methods, such as  
electronic structure simulations, ab-initio molecular dynamics, classical molecular dynamics, 
stochastic methods, lattice kinetic theory, finite difference and finite elements  for continuum fields.   
Each of these methods keeps expanding its range of applicability, to the point of generating 
overlap regions which enable the development of powerful multiscale procedures 
\cite{weinan2011principles,levitt2014birth,yip2013multiscale,broughton1999concurrent,de2011coarse,succi2008lattice,falcucci2011lattice}.

In this paper we focus on a very specific window of material science, 
{\it meso-materials}, namely materials whose ``effective'' constitutive bricks
are neither atoms nor molecules, but droplets instead.

Obviously, droplets cannot  serve as ``super-molecules'' in a literal 
sense, since they generally lack chemical specificity. 
Yet, in recent years, droplets have revealed unsuspected capabilities of serving as 
building blocks (motifs) for the assembly of new types of soft materials, such as
dense foams, soft glasses, microfluidic crystals and many others  \cite{garstecki2006flowing,raven2009microfluidic,abate2009high,beatus2012physics}.

Droplets offer a variety of potential advantages over molecules, primarily the possibility of 
feeding large-scale productions of soft materials, such as scaffolds for tissue engineering 
and other biomedical applications \cite{fernandez2016fluids,piazza2011soft,mezzenga2005understanding}.

From a computational standpoint, mesoscale materials offer the opportunity to deploy mesoscale models
with reduced needs of down-coupling to molecular dynamics and even less for to upward coupling
to continuum methods. Mesoscale particle methods \cite{gompper2009multi,noguchi2007particle} and especially lattice Boltzmann
methods \cite{succi2018lattice,kruger2017lattice,dorazio2003} stand out as major techniques in point.             
In this paper we focus precisely on the recent extensions of the latter which are providing a versatile
tool for the computational study of such droplet-based mesoscale materials. 

The main issue of this class of problems is that they span six or more orders of magnitude in space (nm to mm) 
and nearly twice as many in time, thus making the direct simulation of their dynamics unviable even on 
most powerful present-day supercomputers. 
Why six orders? Simply because the properties of such materials are to a large
extent controlled by nanoscale interactions between near-contact fluid-fluid interfaces, 
which affect the behaviour of the entire device, typically extending
over millimiter-centimeters scales.
Why twice as many in time? 
Typically, because the above processes are driven by capillary forces, resulting in very 
slow motion and long equilibration times, close to the diffusive scaling $t_{dif} \sim L^2$, 
$t_{dif}$ being the diffusive equilibration time of a device of size $L$. 

Notwithstanding such huge computational demand, a clever combination of large-scale computational 
resources and advanced multiscale modelling techniques may provide decisive advances in the 
understanding of the aforementioned complex phenomena and on the ensuing computational design of new mesoscale materials.     
Many multiscale techniques have emerged in the past two decades, based on static or moving grid methods, as well as
various forms of particle dynamics, both at the atomistic and mesoscale levels \cite{feng2004immersed,praprotnik2006adaptive,tiribocchi2019curvature}.  

In this paper, we shall be concerned mostly with a class of mathematical models known as Lattice Boltzmann (LB) 
methods, namely a lattice formulation of a Boltzmann's kinetic equation, which has found widespread use 
across a broad range of problems involving complex states of flowing matter at all scales, 
from macroscopic fluid turbulence, to micro and nanofluidics \cite{succi2018lattice}.

Among others, one of the main perks of the LB method is its excellent amenability to massively
parallel implementations, including forthcoming Exascale platforms.
From the computational point of view an Exascale system (the supercomputer class foreseen 
in 2022) will be able to deliver a peak performance of 
more then one billion of billions floating-point operations per second ($10^{18}$), an impressive figure that, once properly
exploited, can benefit immensely soft matter simulations.

As a matter of fact, attaining Exaflop performance on these applications is an open challenge, as it requires 
a synergic integration of many different skills and a performance-oriented match between system architecture 
and the algorithmic formulation of the mathematical model.

In this paper we provide a description of the different key points that need to be addressed to achieve
Exascale performance in Lattice Boltzmann simulations of soft mesoscale materials.

The article is organized as follows. 

In section \ref{LBE_intro}, for the sake of self-consistency, we provide
a brief introduction to the main features of the LB method.

In section \ref{Exascale} a qualitative description of the up-to-date performance of pre-exascale computers
is discussed, together with an eye at LB performances on such pre-exascale (Petascale) computers 
{\it for macroscopic hydrodynamics}.

In section \ref{par:Microfluidics} a concrete LB application to microfluidic problems is presented.
In section \ref{par:Colloidal} a new high-perfomance lattice Boltzmann code for colloidal flows  is presented.

Finally, in section \ref{par:Conclusions} partial conclusions are drawn together with an outlook of future developments

\section{Basics of Lattice Boltzmann Method} \label{LBE_intro}

The LB method was developed in the late 1980's in a (successful) attempt to  remove the statistical
noise hampering the lattice gas cellular automata (LGCA) approach to fluid dynamic simulations \cite{succi2018lattice}.\\
Over the subsequent three decades, it has witnessed an impressive boost of applications across a remarkably broad spectrum 
of complex flow problems, from fully developed turbulence to relativistic flow all the way down to quark-gluon plasmas \cite{succi2008lattice,succi2015lattice}.\\
The key idea behind LB is to solve a minimal Boltzmann Kinetic Equation (BKE) on a suitable phase-space-time crystal.
 
This means tracing the dynamics of a set of discrete distribution functions (often named \textit{populations}) 
$f_i(\vec{x};t)$, expressing the probability of finding a particle at position $\vec{x}$ and time $t$ with a discrete velocity 
$\vec{v}=\vec{c}_i$.

% ------------------------------------------------------------------------------------------------
\begin{figure*}
\begin{center}
  \includegraphics[width=0.5\linewidth]{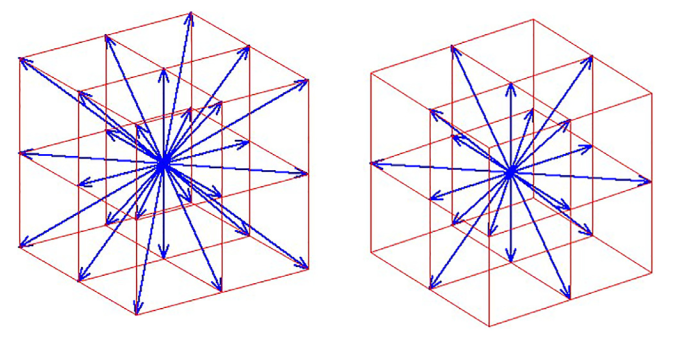}
\end{center}
\caption{Left: D3Q27 lattice, a three dimensional mesh composed of a set of 27 discrete velocities. Right: D3Q19 lattice, a three dimensional mesh composed of a set of 19 discrete velocities.}
\label{fig:D3Q27}
\end{figure*}
% ------------------------------------------------

To correctly solve the Boltzmann kinetic equation, the set of discrete velocities must be chosen in order to secure enough symmetry to 
comply with mass-momentum-energy conservation laws of macroscopic hydrodynamics as well as with rotational symmetry.
In Fig. \ref{fig:D3Q27}, two typical 3D lattices used for current LB simulations are shown, one with a set of 27 velocities (D3Q27, left)
and the other one with  19 discrete velocities (D3Q19, right).

In its simplest and most compact form, the LB equation reads as follows:
\begin{equation} \label{eq:eq1}
f(\vec{x}+\vec{c}_i,t+1) = f'_i(\vec{x};t) \equiv (1-\omega) f_i(\vec{x};t) + \omega f_i^{eq}(\vec{x};t) + S_i,\;\;i=1,b
\end{equation}
where $\vec{x}$ and $\vec{c}_i$ are 3D vectors in ordinary space, $ f_i^{eq} $ are the equilibrium distribution functions and $S_i$ is a source term.
Such equation represents the following situation: the populations at site $\vec{x}$ at time $t$ collides (a.k.a. collision step) and produce a post-collision state $f'_i(\vec{x};t)$,
which is then scattered away to the corresponding neighbour (a.k.a. propagation step) at $\vec{x}_i$ at time $t+1$.

The lattice time step is made unitary, so that $\vec{c}_i$ is the length of the link connecting a generic lattice site node $\vec{x}$ to its $b$ neighbors, located at $\vec{x}_i = \vec{x}+\vec{c}_i$. 
In the D3Q19 lattice, for example,  the index $i$ runs from $1$ to $19$, hence there are $19$ directions of propagation (i.e. neighbors) for each grid-point $\vec{x}$.

The local equilibrium populations are provided by a lattice truncation, to second order in the Mach number $M=u/c_s$, of 
the Maxwell-Boltzmann distribution, namely
\begin{equation} \label{eq:equil}
f_i^{eq}(\vec{x};t) = w_i \rho (1 + u_i + q_i)
\end{equation}
where $w_i$ is a set of weights normalized to unity, $u_i = \frac{\vec{u} \cdot \vec{c}_i}{c_s^2}$ and  $q_{i}=(c_{ia}c_{ib}-c_{s}^{2}\delta_{ab})u_{a}u_{b}/2c_{s}^{4}$, with $c_s$ equal to the speed of sound in the lattice, and an implied sum over repeated latin indices $a,b=x,y,z$.

The source term $S_i$ of Eq.~\ref{eq:eq1} typically accounts for the momentum exchange between the fluid and external (or internal) fields, 
such as gravity or self-consistent forces describing potential energy interactions within the fluid.

By defining fluid density and velocity as
\begin{equation} \label{eq:eq2}
\rho = \sum_i f_i  \hspace{2cm}
\vec{u}=(\sum_i f_i \vec{c}_i)/\rho,
\end{equation}
the Navier-Stokes equations for an isothermal quasi-incompressible fluid can be recovered in the continuum limit if the lattice has the suitable symmetries aforementioned
and the local equilibria are chosen according to Eq.~\ref{eq:equil}.

Finally, the relaxation parameter $\omega$ in Eq.~\ref{eq:eq1} controls the viscosity of the lattice fluid according to
\begin{equation} \label{eq:eq3}
\nu = c_s^2 (\omega^{-1}-1/2).
\end{equation}
Further details about the method can be found in Ref.~\cite{kruger2017lattice,montessori2018lattice}.

One of the main strengths of the LB scheme is that, unlike advection, streaming is i) exact, since it occurs along
straight lines defined by the lattice velocity vectors $\vec{c}_i$, regardless of the complex structure of the fluid flow,  and ii) it is
implemented via a memory-shift without any floating-point operation. 
This also allows to handle fairly complex boundary conditions \cite{leclaire2017generalized} in a more {\it conceptually}
transparent way with respect to other mesoscale simulation techniques \cite{schiller2018mesoscopic}.

\subsection{Multi-component flows}

The LB method successfully extends to the case of multi-component and multi-phase fluids. In a binary 
fluid, for example,  each component (denoted by  $r$ and $b$ for red and blue, respectively) comes with 
its own populations plus a term modeling the interactions between fluids. 

In this case the equations of motion read as follows:
\begin{eqnarray}\label{eq:twobgk}
f^r_{i}(\vec{x}+\vec{c}_{i};t+1)=(1-\omega_{eff})f^r_{i}(\vec{x};t)+\omega_{eff} f_{i}^{eq,r}(\rho^r;\vec{u})+S^r_{i}(\vec{x};t)
\\	
f^b_{i}(\vec{x}+\vec{c}_{i};t+1)=(1-\omega_{eff})f^b_{i}(\vec{x};t)+\omega_{eff} f_{i}^{eq,b}(\rho^b;\vec{u})+S^b_{i}(\vec{x};t)
\\
\omega_{eff}=2c_s^2/(2 \bar{\nu} -c_s^2)
\\
\frac{1}{\bar{\nu}}=\frac{\rho_k}{(\rho_k+\rho_{\bar{k}})}\frac{1}{\nu_k} + \frac{\rho_{\bar{k}}}{(\rho_k+\rho_{\bar{k}})}\frac{1}{\nu_{\bar{k}}}
\end{eqnarray}
where $\omega_{eff}$ is related to the kinematic viscosity $\bar{\nu}$ of the mixture of the two fluids.
\\

The extra term $S^r_{i}(\vec{x};t)$ of Eq.~\ref{eq:twobgk} (and similarly for $S^b$) can be computed as the difference 
between the local equilibrium population, calculated at a shifted fluid velocity, and the one taken at the 
effective velocity of the mixture \cite{kupershtokh2004new}, namely:
\begin{equation}
S^r_{i}(\vec{r};t)=f_{i}^{eq,r}(\rho^r,\vec{u}+\frac{\vec{F}^{r} \Delta t}{\rho^r})-f_{i}^{eq,r}(\rho^r,\vec{u}).\label{eq:edm}
\end{equation}

Here $\vec{F}^r$ is an extra cohesive force, usually defined as \cite{shan1993lattice} 
\begin{equation}\label{shanchen_force}
\vec{F}^{r}(\vec{x},t)=\rho^{r}(\vec{x},t)G_C\sum_{i}w_{i}\rho^{b}(\vec{x}+\vec{c}_{i},t)\vec{c_{i}},
\end{equation}
capturing the interaction between the two fluid components.
In Eq. \ref{shanchen_force}, $G_C$ is a parameter tuning the strength of this intercomponent force, and takes positive values
for repulsive interactions and negative for attractive ones.

This formalism has proved extremely valuable for the simulation of a broad variety of multiphase and multi-component fluids and represents a major mainstream of current LB research.

\section{LB on Exascale class computers} \label{Exascale}

Exascale computers are the next major step in the high performance computing arena, the 
deployment of the first Exascale class computers being planned at the moment for 2022. 
To break the barrier of $10^{18}$ floating point operations per second, this class of machines will be based on 
a hierarchical structure of thousands of nodes, each with up to hundreds cores using many accelerators 
like GPGPU (General Purpose GPU or similar devices) per node. 
Indeed, a CPU-only Exascale Computer is not feasible due to heat dissipation constraints, as it 
would demand more than of 100MW of electric power!

As a reference, the current top-performing Supercomputer (Summit) in 2019 is composed of 
4608 nodes, with 44 cores per node and 6 Nvidia V100 GPU per node,
falling short of the Exascale target by a factor five, namely 200 Petaflops (see online at Top500, \cite{top500}), 
and over 90\% of the computational performance being delivered by the GPUs.\\

Since no major improvement of single-core clock time is planned, due again to heat 
power constraints, crucial 
to achieve exascale performance is the ability to support concurrent execution of many tasks in parallel, 
from  $O(10^4)$ for hybrid parallelization (e.g. OpenMP+MPI or OpenAcc+MPI) up to $ O(10^7) $ for a 
pure MPI parallelization.\\

Hence, three different levels of \textit{parallelism} have to be implemented for an efficient Exascale simulation:  
i) at core/floating point unit level: (Instruction level parallelism, vectorization),
ii) at node level: (i.e., shared memory parallelization with CPU or CUDA, for GPU, threads),
iii) at cluster level: (i.e., distributed memory parallelization among tasks).

All these levels have to be efficiently orcherstrated to achieve performance by the 
user's implementations, together with the tools (compilers, libraries) and technology 
and the topology of the network, as well as the efficiency of the communication software.
Not to mention other important issues, like reliability, requiring a fault 
tolerant simulation environment \cite{da2015exascale,snir2014addressing}.\\

How does LB score in this prospective scenario?\\

In this respect, the main LB features are as follows:
\begin{itemize}
	\item Streaming step is exact (zero round-off)
	\item Collision step is completely local (zero communication)
	\item First-neighbors communication (eventually second for high-order formulations)
	\item Conceptually easy-to-manage boundary conditions (e.g. porous media)
        \item Both pressure and fluid stress tensor are locally available in space and time  
        \item Emergent interfaces (no front-tracking) for multi-phase/species simulation
\end{itemize}
All features above are expected to facilitate exascale implementations \cite{succi2019towards}. In Ref.~\cite{liu2019sunwaylb}, an overview of different LB code implementations on a variety of large-scale HPC machines is presented, and it is shown that LB is, as a matter of fact, in a remarkable good position to exploit Exascale systems.\\ 

In the following we shall present some figures that can be reached with LB codes on exascale systems. \\

\begin{figure*} \begin{center}
\includegraphics[width=0.7\linewidth]{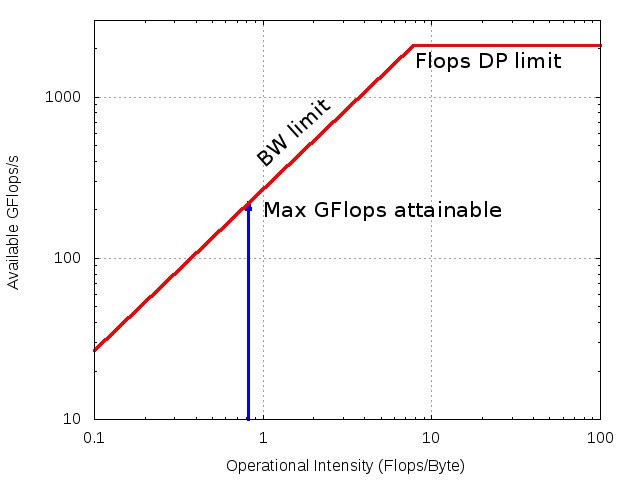}
\end{center}
\caption{Roofline model for single-phase, single time relaxation, Lattice Boltzmann. 
The vertical line indicates the performance range using double precision.}
\label{fig:roof}
\end{figure*}

According to the roofline model \cite{williams2009roofline}, achievable performance can be ranked in terms of Operational Intensity (OI), defined as the ratio between flops performed and data that need to be loaded/stored from/to memory. 
At low OI ($ < 10 $), the performance is limited by the memory bandwidth, while for higher values  the limitation comes from the floating point units availability. 
It is well known that LB is a bandwidth limited numerical scheme, like any other CFD model. Indeed, the OI index for LB schemes is around 0.7 for double precision (DP) simulations using a D3Q19 
lattice\footnote{For a single fluid, if $F  \simeq 200 \div 250 $ is the number of floating point operations per lattice site and time step and $B= 19 \times 2 \times 8=304 $ is load/store demand in bytes (using double precision), the operational intensity is $F/B \sim 0.7$. 
For Single-precision simulation is 1.4} (see Fig.~\ref{fig:roof} where the roofline 
for LB is shown\footnote{Bandwidth and Float point computation limits are obtained 
performing {\em stream} and HPL benchmark.}).\\

\begin{figure*}
\begin{center}
\includegraphics[width=0.7\linewidth]{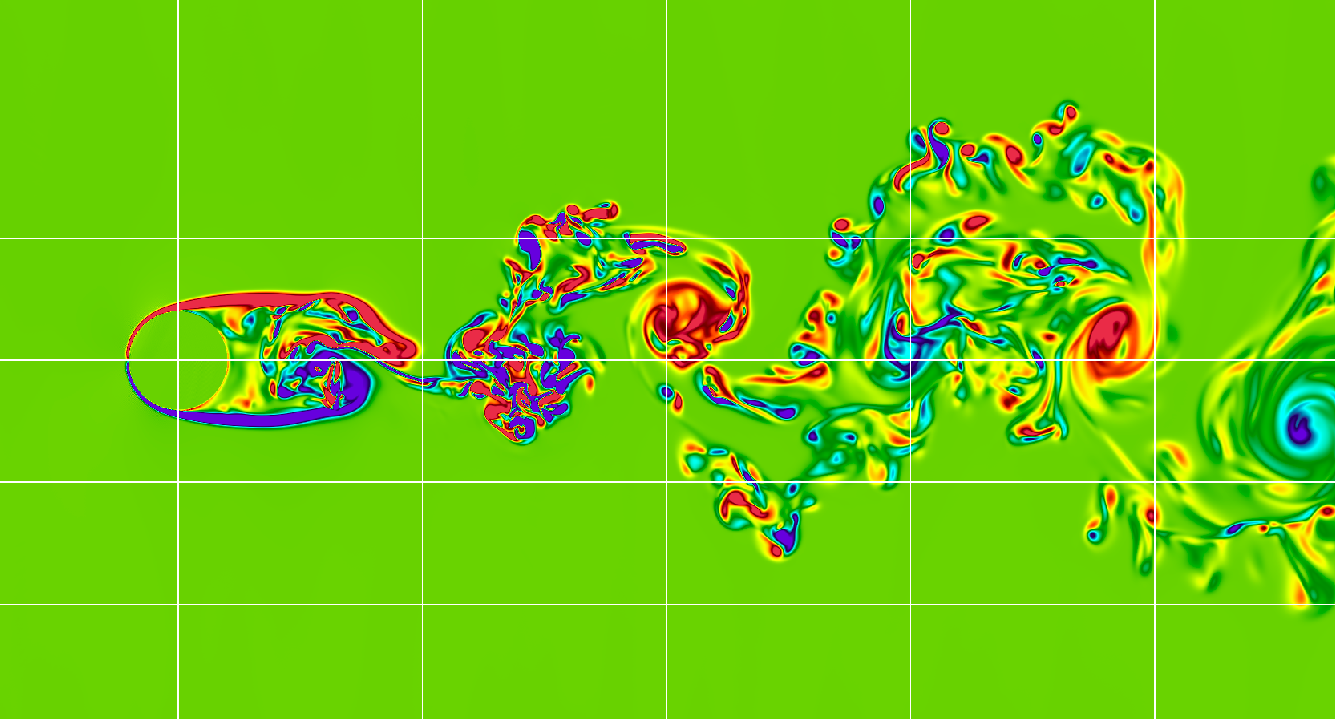}
\end{center}
\caption{Vorticity of a flow around a cylinder at Reynolds number $=2000$. 
The color map indicates the direction of rotation, blue for clockwise and 
red for counterclockwise. The flow was simulated with an optimised 
LB code for macroscopic hydrodynamics \cite{succi2019towards}}
\label{fig:CYL}
\end{figure*}

In Fig.~\ref{fig:CYL}, a 2D snapshot of the vorticity of the flow around a 3D cylinder at $ Re=2000 $ 
is shown (\cite{DSFD2020} in preparation).\\ 
This petascale class simulation was performed with an hybrid MPI-OpenMP parallelization 
(using 128 Tasks and 64 threads per task), using a single-phase, single time 
relaxation, 3D lattice. Using $O(10^4)$ present-day GPUs\footnote{At this time only GPUs seems the only device mature enough to be used for a Exascale Machine.}, a hundred billion lattice simulation would complete one million time-steps in something between $1$ and $3$ hours, corresponding to about 3 and 10 TLUPS (1 Tera LUPS is a trillion Lattice Units per Second). So using an Exascale machine  more realistic structure, both in terms of size, complexity (i.e. decorated structure) and simulated time can be performed \cite{AmatiFalcucci2018}. 
To achieve this we must be able to handle order of $ 10^4 $ tasks, and each of them 
must be split in many, order of $1'000 $, GPUS threads-like processes. \\ 
% ------------------------------ Fig 1 
% ---------------------------- Fig 2

What does this mean in terms of the multiscale problems sketched in the opening of this article?

With a one micron lattice spacing and one nanosecond timestep, this means simulating
a cubic box 5 mm in side over one millisecond in time.
Although this does not cover the full six orders in space, from, say, 10 nanometers to centimeters, which characterize 
most meso-materials, it offers nonetheless a very valuable order of magnitude boost in size 
as compared to current applications.  

\section{LB method for microfluidic crystals}   \label{par:Microfluidics}

As mentioned in the Introduction, many soft matter systems host concurrent interactions 
encompassing six or more decades in space and  nearly twice as many in time.
Two major directions can be endorsed to face this situation: the first consists 
in developing sophisticated multiscale methods capable of covering five-six spatial decades 
through a clever combination of advanced computational techniques, such as local 
grid-refinement, adaptive grids, or grid-particle combinations \cite{mehl2019adaptive,lahnert2016towards,lagrava2012advances,dupuis2003theory,filippova1998grid}.
 
The second avenue consists in developing suitable coarse-grained models, operating 
at the mesoscale, say microns, through the incorporation of effective forces and potentials designed 
in such a way as to retain the essential effects of the fine-grain scales on the coarse-grained ones 
(often providing dramatic computational savings)

Of course, the two strategies are not mutually exclusive; on the contrary, they should be combined
in a synergistic fashion, typically employing as much coarse-graining as possible to reduce the need
for high-resolution techniques.

In \cite{montessori2019mesoscale}, the second strategy has been successfully adopted to the simulation of microdevices. 
Experiments  \cite{raven2009microfluidic,marmottant2009microfluidics}  have shown that a soft flowing microfluidic crystal can 
be designed by air dispersion in the fluid with a flow focuser: the formation of drops is due to 
the balance between pressure drop, due to the sudden expansion of the channel, and the shear stress, exerted 
by the continuous phase inside the nozzle.

\begin{figure*}
\begin{center}
\includegraphics[width=.45\linewidth]{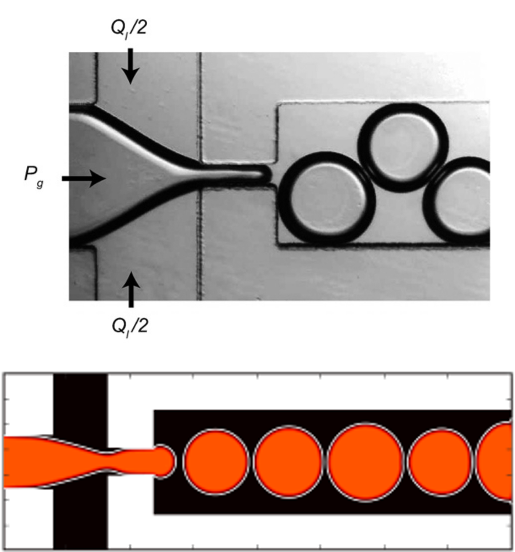}
\end{center}
\caption{(Top). Flow focuser used for the production of air bubbles \cite{raven2009microfluidic,marmottant2009microfluidics}. 
Gas is injected from the horizontal branch at pressure $P_g$ while air enters at flow rate $Q_l/2$ from the two vertical braches. 
They are focused into a striction of width $\simeq 100$ $\mu$m, and resulting air bubbles are collected downstream in the outlet chamber. (Bottom) Lattice Boltzmann simulation showing the production of roughly mono-disperse fluid droplets within a microfluidic flow focuser. The red phase represents the dispersed fluid (oil) while the black phase the continuous one (water).}
\label{fig:micro1}        
\end{figure*}

In Fig. \ref{fig:micro1} (top), we show the typical experimental setup for the production of 
ordered dispersion of mono-disperse air droplets. 

% ------------------------------  Fig 3
In our LB experiments (Fig.~\ref{fig:micro1}, bottom), droplet formation is controlled by tuning i) the dispersion-to-continuous flow ratio $\alpha$ (defined as $\alpha=u_d/2u_c$, where $u_d$ and $u_c$ are the speeds of the dispersed and the continuous phase at the inlet channel) and by ii) the mesoscopic force $F_{rep}$ modeling the repulsive effect of a surfactant confined at the fluid interface.

\begin{figure}[h] \label{fig:micro0}
\begin{center}
\centerline{\resizebox{.4\linewidth}{!}{\includegraphics{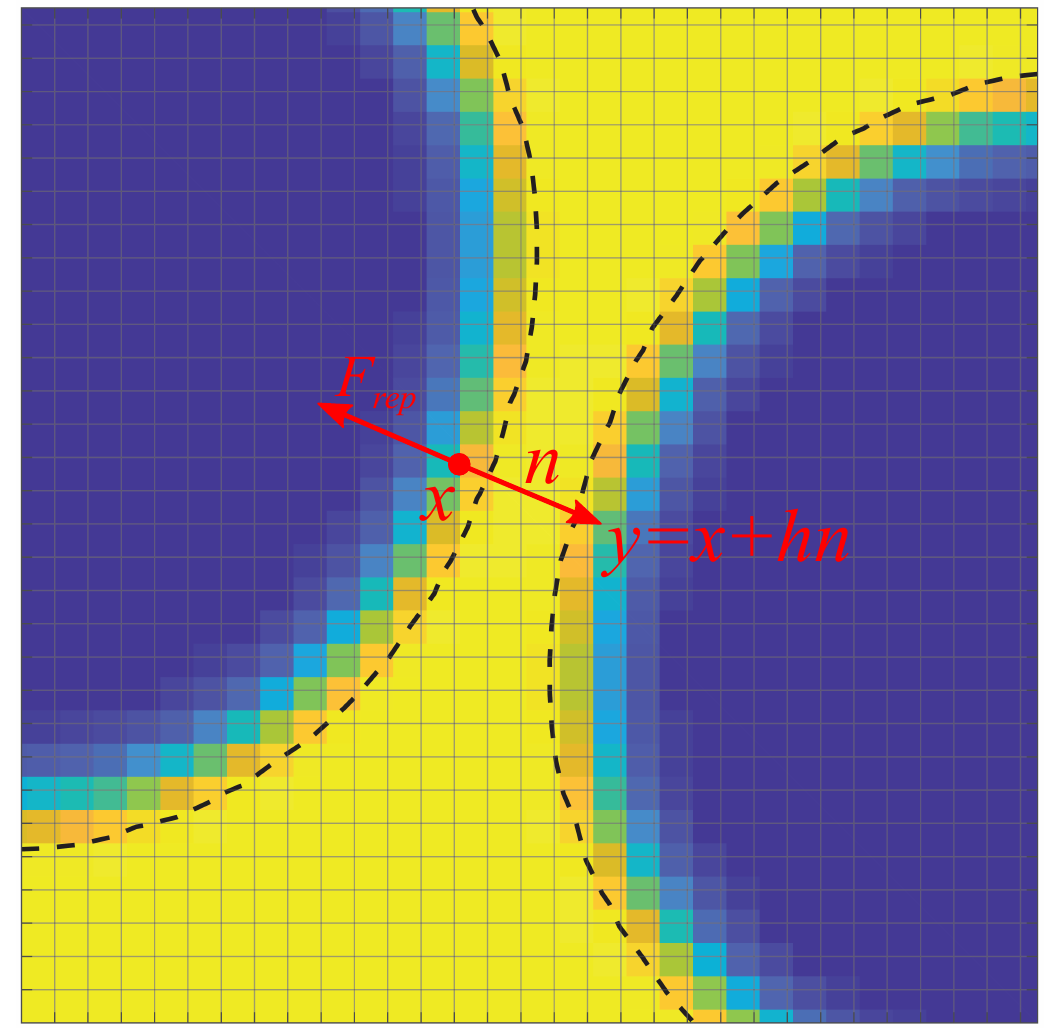}}}
\end{center}
\caption{Close-up view of the modelling of the near interaction between two droplets. 
$F_{rep}$ is the repulsive force and $n$ is the unit vector perpendicular to the interfaces, 
while $x$ and $y$ indicate the positions, at distance $h$, located within the fluid interface.
}
\end{figure}

The dispersed phase (in red in Fig.~\ref{fig:micro1}, bottom) is pumped with a predefined speed $u_d$ within the 
horizontal branch, whereas the continuous phase (black) comes from the two vertical branches at speed $u_c$. 
They are driven into the orifice ,where the droplet form and are finally collected in the outlet chamber.
A schematic representation of $F_{rep}$ on the lattice is reported in Fig.\ref{fig:micro0}.
This term enters the LB equation as a forcing contribution acting solely at the fluid interfaces when in close contact. 
Its analytical expression is
\begin{equation}
F_{rep} = -A_h [h(x)] n \delta_I,
\end{equation}
where $\delta_I\propto \nabla\psi$ is a function, proportional to the fluid concentration $\psi$, confining the near-contact force at the fluid interface, while
$A_h[h(x)]$ sets the strength of the near-contact interactions. It is equal to a positive constant $A$ if $h<h_{min}$ and it decays as $h^{-3}$ if $h>h_{min}$, with
$h_{min}=3-4$ lattice units. 
Although other functional forms of $A[h(x)]$ are certainly possible, this one proves sufficient to capture 
the effects occurring at the sub-micron scale, such as the stabilization of the fluid film formed between 
the interfaces and the inhibition of droplet merging.

% -------------------------------------------   Fig 4

An appropriate dimensionless number capturing the competition between surface 
tension $\sigma$ and the near-contact forces $F_{rep}$ can be defined as 
$N_c = A \Delta{x} / \sigma$, where $\Delta{x}$ is the lattice spacing. 

Usually, if $N_c \ll 1$, capillary effects dominate and drops merge, whereas if $N_c \sim 1$, close 
contact interactions prevail and droplet fusion is inhibited.
A typical arrangement reproducing the latter case is shown in Fig. \ref{fig:micro1}, obtained for $A_h = 1$ and $N_c = 0.1$.

These results suggest that satisfactory compliance with experimental results can be achieved by means of 
suitable coarse-grained models, which offer dramatical computational savings over grid-refinement methods.

However, success or failure of coarse graining must be tested case by case, since approximations 
which hold for some materials, say dilute microfluidic crystals, may not necessarily apply 
to other materials, say dense emulsions in which all the droplets are
in near-touch with their neighbours.
In this respect, significant progress might be possible by resorting to Machine-Learning techniques 
\cite{lecun2015deep,goodfellow2016deep}, the idea being of semi-automating the procedure of developing customized
coarse-grained models, as detailed in the next section. 

\subsection{Machine-learning for LB microfluidics}

Machine learning has taken modern science and society by storm. 
Even discounting bombastic claims mostly devoid of scientific value, the fact remains that the idea of 
automating difficult tasks through the  aid of properly trained neural networks may add a 
new dimension to the space of scientific investigation \cite{CARLEO,BRENNER,SINAI}. 
For a recent critical review, see for instance \cite{BIGDATA}.

In the following, we portray a prospective machine-assisted procedure to facilitate the computational design of
microfluidic devices for soft mesoscale materials.
The idea is to ``learn'' the most suitable coarse-grained expression of the Korteweg tensor, the
crucial quantity controlling non-ideal interactions in soft mesoscale materials.

The procedure develops through three basic steps: 
1) Generate high-resolution data via direct microscale simulations; 
2) Generate coarse-grained data upon projection (averaging) of high-resolution data;    
3) Derive the coarse-grained Korteweg tensor using machine learning techniques fed with data from step 2).

The first step consists of performing very high resolution simulations 
of a microfluidic device, delivering the fluid density $\rho_f(\vec{x}_f,t_f)$, the flow field $\vec{u}_f(\vec{x}_f,t_f)$ at each lattice location 
$\vec{x}_f$ and given time instant $t_f$ of a fine-grid simulation with $N_f$ lattice sites and $M_f$
timesteps, for a total of $D_f= 4 N_f M_f$ degrees of freedom in three spatial dimensions.  
To be noted that such fine-grain information may be the result of an underlying molecular
dynamics simulation. 

The second step consists of coarse-graining the high-resolution data to 
generate the corresponding ``exact'' coarse-grained data.
Upon suitable projection, for instance averaging over blocks of $B=b^4$ 
fine-grain variables, $b$ being the spacetime blocking factor, this provides the
corresponding (``exact'') coarse-grained density and velocity 
$\rho_c$ and $\vec{u}_c$, for a total of $D_c=D_f/B \ll D_f$ degrees of freedom.

The third step is to devise a suitable model for the Korteweg tensor at 
the coarse scale $x_c = b x_f$. 
A possible procedure is to postulate parametric expressions of the coarse-grained
Korteweg tensor, run the coarse grained simulations and perform a systematic search
in parameter space, by varying the parameters in such a way as to minimize the departure between
the "exact" expression $K_c = \mathcal{P} \rho_f$ obtained by projection of the 
fine-grain simulations and the parametric expression $K_c[\rho_c;\lambda]$, 
where $\lambda$ denotes the parameters of the coarse-grained model, typically 
the amplitude and range of the coarse-grained forces. 

The optimization problem reads as follows: 
find $\lambda$ such as to minimise the error,
\begin{equation}
e[\lambda] =  \left\lVert (P_c, P_c[\lambda]) \right\rVert
\end{equation}
where  $ \left\lVert ..\right\rVert$ denotes a suitable metrics in 
the $D_c$-dimensional functional space of solutions. 

This is a classical and potentially expensive optimization procedure.

A possible way to reduce its complexity is to leverage the machine 
learning paradigm by instructing a suitably designed neural network 
(NN) to ``learn'' the expression of $K_c$ as a functional of the 
coarse-grained density field $\rho_c$.
Formally:
\begin{equation}
\label{ML}
K_c^{ml} = \sigma_L [W \rho_c]
\end{equation}
where $\rho_c = \mathcal{P} \rho_f$ is the ``exact'' coarse-grained density and  
$\sigma_L$ denotes the set of activation functions of a $L$-levels deep neural 
network, with weights $W$.

Note that the left hand side is an array of  $6 D_c$ values, six being the number of independent components
of the Korteweg tensor in three dimensions, while the input array $\rho_c$ contains only $D_c$ entries.
Hence, the set of weights in a fully connected NN contains $36 N_c ^2$ entries.  
This looks utterly unfeasible, until one recalls that the Korteweg tensor 
only involves the Laplacian and gradient product of the density field 
$$
K_{ab}=\lambda [\rho \Delta \rho+\frac{1}{2}(\nabla \rho)^2] \delta_{ab} 
      -\lambda \nabla_a \rho \nabla_b \rho
$$ 
where $a,b=x,y,z$ and $\lambda$ controls the surface tension.

The coarse-grain tensor is likely to expose additional nonlocal terms, but 
certainly no global dependence,
meaning that the set of weights connecting the value of the $K$-tensor 
at a given lattice site to the density field should be of the order of O(100) at most. 

For the sake of generality, one may wish to express it in integral form
\begin{equation}
K_c(\vec{x}_c)= \rho(\vec{x}_c) \int G(\vec{x}_c,\vec{y}_c) \rho(\vec{y}_c) d \vec{y}_c
\end{equation}
and instruct the machine to learn the kernel $G$.

This is still a huge learning task, but not an unfeasible one, the number of 
parameters being comparable to similar efforts in ab-initio molecular dynamics, whereby
the machine learns multi-parameter coarse-grained potentials \cite{CAR,PAR}.

Further speed can be gained by postulating the functional expression of 
the coarse grained $K$-tensor in formal analogy with recent work on 
turbulence modelling \cite{RUI}, i.e. based on the basic symmetries
of the problem, which further constrains the functional dependence 
of $K_c$ on the density field.

Work is currently in progress to implement the aforementioned ideas.

\section{High-performance LB code for bijel materials} \label{par:Colloidal}

For purely illustrative puroposes,
in this section we discuss a different class of complex flowing systems made up 
of colloidal particles suspended in a binary fluid mixture, such as oil and water.

A notable example of such materials is offered by {\it bijels} \cite{stratford2005colloidal}, soft materials 
consisting of a pair of bi-continuous fluid domains, frozen into a permanent porous
matrix by a densely monolayer of colloidal particles adsorbed onto the fluid-fluid interface. 

The mechanical properties of such materials, such as elasticity and pore size, can be fine-tuned 
through the radius and the volume fraction of the particles, typically
in the range $0.01 < \phi < 0.1$, corresponding to a number
$$
N = \frac{3 \phi}{4 \pi} (\frac{L}{R})^3 
$$
of colloids of radius $R$ in a box of volume $L^3$.

Since the key mechanisms for arrested coarsening are i) the sequestration of the 
colloids around the interface and ii) the replacement of the interface by the colloid 
itself, the colloidal radius should be significantly larger than
the interface width, $R/w >2$. 
Given that LB is a diffuse-interface method and the interfaces span a few 
lattice spacings (say three to five), the spatial hierarchy for a typical large scale simulation 
with, say, one billion grid points reads as follows:    
$$
dx=1,\; w=3 \div 5, \; R=10 \div 50,  L = 1000
$$
With a typical volume fraction $\phi=0.1$, these parameters correspond to about $N \sim 10^3 \div 10^5$ colloids.
 
In order to model bijels, colloidal particles are represented as rigid spheres of 
radius $R$, moving under the effect
of a total force ($\vec{F}_p$) and torque ($\vec{T}_p$), both acting on the 
center of mass $\vec{r}_p$ of the $p$-th particle.

The particle dynamics obeys Newton’s equations of motion (EOM):
\begin{equation}\label{newton}
\frac{d \vec{r}_p}{d t}=\vec{u}_p,\hspace{0.5cm} m_p \frac{d \vec{u}_p}{d t}=\vec{F}_p,\hspace{0.5cm}I_p \frac{d \vec{\omega}_p}{d t}=\vec{T}_p, 
\end{equation}
where $\vec{u}_p$ is the particle velocity and $\vec{\omega}_p$ is the corresponding angular velocity. 
The force on each particle takes
into account the interactions with the fluid and the inter-particle forces, 
including the lubrication term \cite{ladd1994numerical,nguyen2002lubrication}.
Hence, the equations of motion (EOM) are solved by a leapfrog approach, 
using quaternion algebra for the rotational component (\cite{rozmanov2010robust,svanberg1997research}).

The computation is parallelised using MPI. While for the hydrodynamic part 
the domain is equally distributed (according to a either 1D, or 2D or 3D domain decomposition), 
for the particles a complete replication of their physical coordinates 
(position, velocity, angular velocity) is performed. 

More specifically, $\vec{r}_p$, $\vec{v}_p$ and $\vec{\omega}_p$ are allocated 
and maintained in all MPI tasks, but 
each MPI task solves EOM only for the particles whose centers of mass 
lie in its sub-domain, defined by the fluid partition. 
Whenever a particle crosses two or more sub-domains, the force and torque are computed with 
an MPI reduction and once the time step is completed, new values of 
$\vec{r}_p$, $\vec{v}_p$ and $\vec{\omega}_p$ are broadcasted to all tasks.

\begin{figure}[h]
\begin{center}
\includegraphics[width=1.0\linewidth]{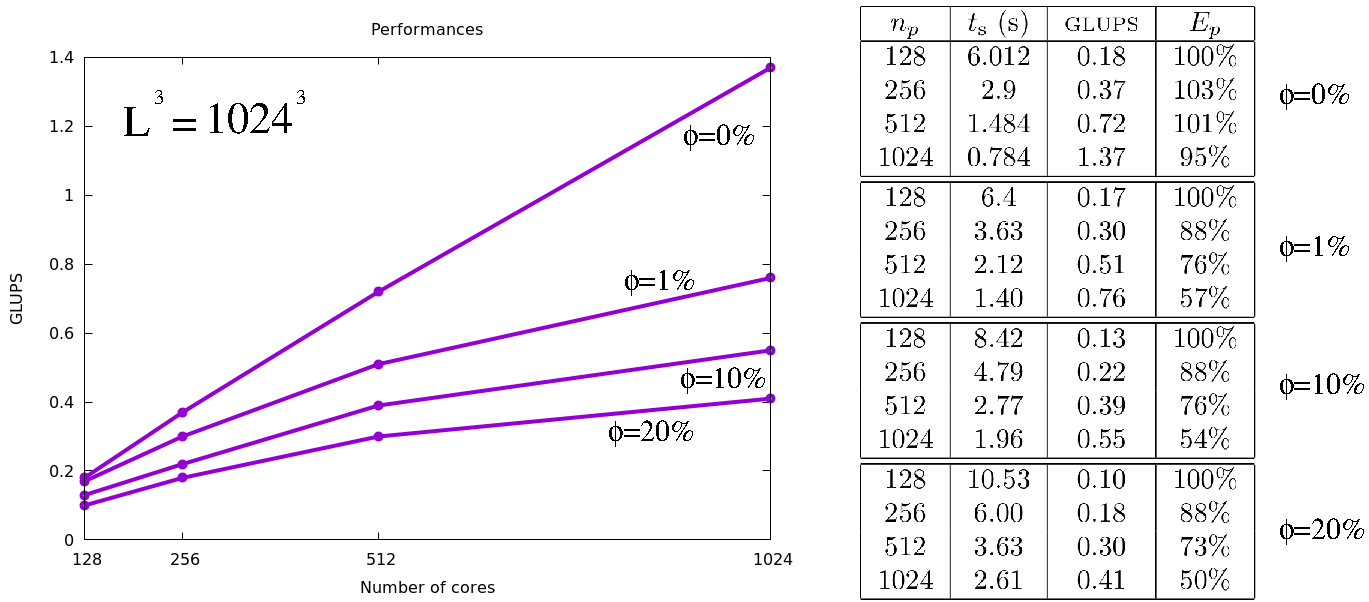}
%\centerline{\resizebox{1.0\linewidth}{!}{\includegraphics{fig6.png}}}
\end{center}
\caption{Left: GLUPS versus number of cores measured in a cubic box of linear size $L=1024$, in which spherical colloids are dispersed in a bicontinuous fluid. From the top to the bottom, the box was filled with $N=0$, $N=15407$, $N=154072$, and $N=308144$ colloids, corresponding to a particle volume fraction $\phi$ equal to 0\%, 1\%, 10\% and 20\%, respectively. Right: Run (wall-clock) time, in seconds, per single time step iteration, $t_{\text{s}}$, with corresponding GLUPS and parallel efficiency, $E_p$, versus the number of computing cores, $n_p$ for the same system. Note that the parallel efficiency is reported in percentage.}
\label{fig6}
\end{figure}

Even at sizeable volume fractions (such as $\phi=0.2$), the number of particles is of the order of 
ten-twenty thousands, hence much smaller than the dimension of the simulation box ($L^3=1024^3$), which is
why the ``replicated data'' strategy (\cite{smith1991molecular,smith1993molecular}) does not significantly 
affect the MPI communication time (see Fig.\ref{fig6}).

%--------------------------------------------  Fig 

% ---------------------------------------------------------------------------------------------------------------
\begin{figure*}
\begin{center}
%\centerline{\resizebox{.45\linewidth}{!}{\includegraphics{fig7.png}}}
\includegraphics[width=.45\linewidth]{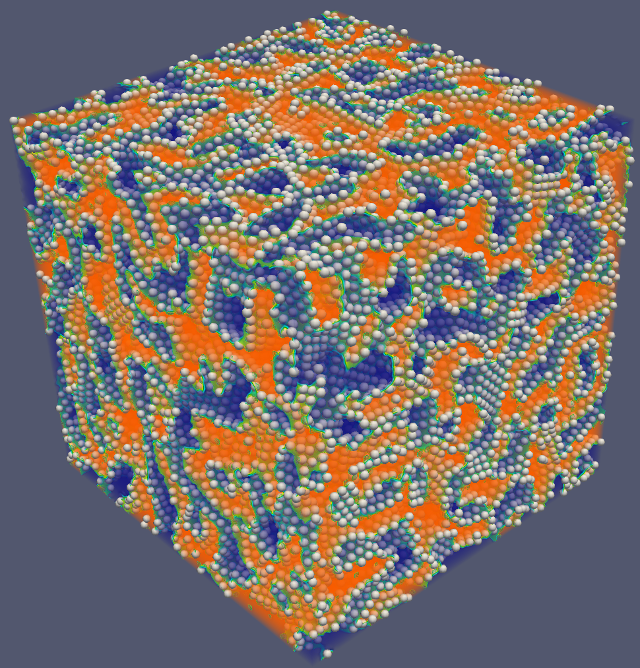}
\end{center}
\caption{Typical morphology of a bijel material obtained by dispersing rigid colloids (of radius $R=5.5$~lu) in a bicontinuous fluid. 
Particles are adsorbed at the fluid interface and, if their volume fraction is sufficiently high, domain coarsening arrests. Red and blue colors indicate the two fluid components while grey spheres are colloids.}
\label{FigLadd}
\end{figure*}
\begin{figure}[h]
	\begin{center}
	\includegraphics[width=0.5\linewidth]{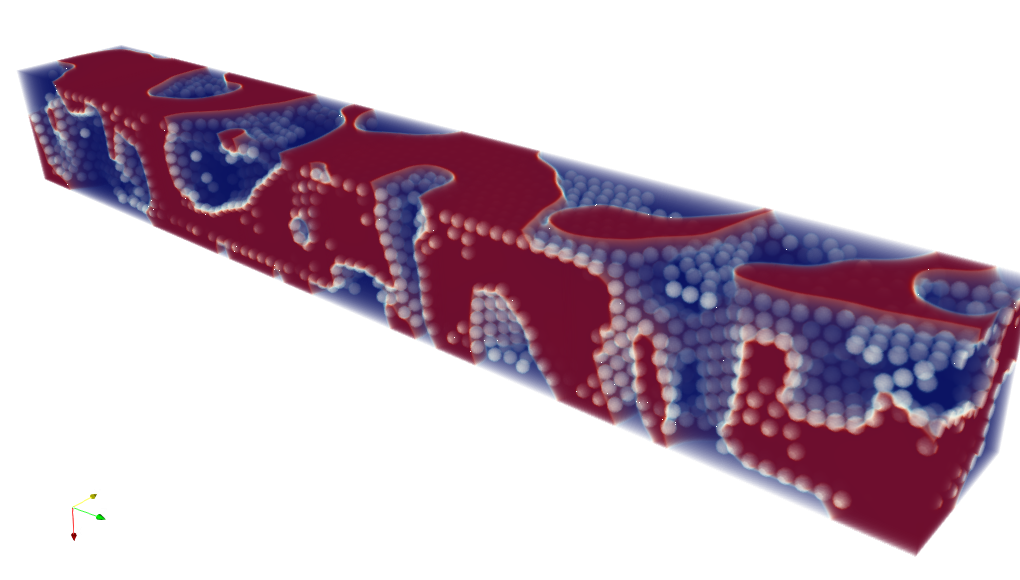}
	\end{center}
	\begin{center}
		\includegraphics[width=0.5\linewidth]{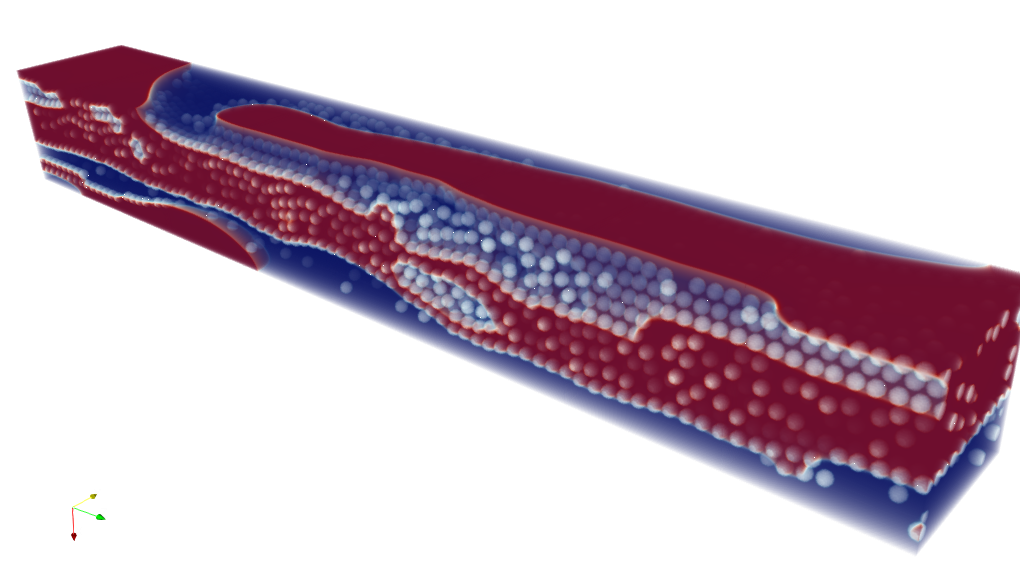}
	\end{center}
	\caption{Bijel in a channel of $128 \times 128 \times 1024$ lu: 
top and bottom walls slide with opposite 
directions (aligned with the mainstream axis $z$). 
Left-right walls are bounce-back while the mainstream axis $z$ is periodic. 
The main parameters are as follows: wall speed $U=0.01$, volume fraction $\phi=0.15$, sphere radius $R=5.5$ lu. 
Top: Initial configuration. Bottom: Elongated structures after half-million steps with applied shear.}
	\label{FigShearBijel}
\end{figure}

% ------------------------------------------------
Results from a typical simulation of a bijel are shown in Fig.\ref{FigLadd}, in which solid 
particles accumulate at the interface leading to the arrest of the domain 
coarsening of the bi-continuous fluid, which in turn supports the
the formation of a soft and highly porous fluid matrix.
Further results are shown in Fig \ref{FigShearBijel}, in which the bijel is confined within solid walls 
with opposite speeds $\pm U = 0.01$, hence subject to a shear $S=2U/H$, $H$ being the channel width.
The complex and rich rheology of such confined bijels is a completely open 
research item, which we are currently pursuing through extensive simulations 
using LBsoft, a open-source software for soft glassy emulsion simulations \cite{bonaccorso2020lbsoft}.

Figure \ref{FigShearBijel} reports the bijel before and after applying the shear 
over half-million timesteps. From this figure, it is apparent that the shear 
breaks the isotropy of the bijel, leading to a highly directional structure
aligned along the shear direction.
Further simulations show that upon releasing the shear for 
another half-million steps does not recover the starting condition, showing 
evidence of hysteresis.

This suggests the possibility of controlling the final shape of the bijel 
by properly fine-tuning the magnitude of the applied shear, thereby opening 
the possibility to exploit the shear to imprint the desired shape 
to bijels within controlled manufacturing processes.

Systematic work along these lines is currently in progress

As to performance, LBsoft delivers GLUPS on one-billion gridpoint configurations 
on large-scale parallel platforms, with a
parallel efficiency ranging from above ninety percent for plain LB (no colloids), down 
to about fifty percent with twenty percent colloidal volume fractions (see Fig.~\ref{fig6}).

Assuming parallel performance can be preserved up to TLUPS, on a Exascale computer, one 
could run a trillion gridpoint simulation (four decades in space) over 
one million timesteps in about two weeks wall-clock time. 

Setting the lattice spacing at $1$ nm to fully resolve the colloidal diameter ($10-100$ nm), this 
corresponds to a sample of material of $10$ micron in side, over a time span of about one microsecond.

By leveraging dynamic grid refinement around the interface, both space and time 
spans could be boosted by two orders of magnitude.
However, the programming burden appears fairly significant, especially due to the presence 
of the colloidal particles. 

Work along these lines is also in progress.

\newpage

\section{Summary and outlook} \label{par:Conclusions}

Summarising, we have discussed some of the main challenges and prospects 
of Exascale LB simulations of soft flowing systems for the design of novel 
soft mesoscale materials, such as microfluidic crystals and colloidal bijels.

Despite major differences in the basic physics, both systems raise a major 
challenge to computer modelling, due to the
coexistence of dynamic interactions over about six decades in space, from 
tens of nanometers for near-contact interactions,
up to the centimeter scale of the experimental device. 
Covering six spatial decades by direct numerical simulations is beyond
reach even for Exascale computers, which permit to span basically four (a trillion gridpoints).
The remaining two decades can either be simulated via local-refinement 
methods, such as multigrid or grid-particle hybrid
formulations, or by coarse-graining, i.e. subgrid modeling of the near-contact 
interactions acting below the micron scale.
Examples of both strategies have been discussed and commented on.

We conclude with two basic takehome's: 1) Extracting exascale performance 
from exascale computers requires a concerted multi-parallel approach, 
no ``magic paradigm'' is available, 2) even with exascale performance
secured, only four spatial decades can be directly simulated, hence 
many problems in soft matter research will still require an 
appropriate blend of grid-refinement techniques and coarse-grained models.
The optimal combination of such two strategies is most likely 
problem-dependent, and in some fortunate instances, the latter alone may suffice.
However, in general, future generations of computational soft matter scientists should be 
prepared to imaginative and efficient ways of combining the two.

\section{Acknowledgements}

S. S., F. B., M. L., A. M. and A. T. acknowledge funding from the European Research Council under the European Union's Horizon 2020 Framework Programme (No. FP/2014-2020)
ERC Grant Agreement No.739964 (COPMAT).

\bibliographystyle{elsarticle-num}
\bibliography{jcs_final}

\end{document}